\documentclass[journal,twoside,web]{ieeecolor}

\usepackage{tmi}
\usepackage{cite}
\usepackage{amsmath,amssymb,amsfonts}
\usepackage{graphicx}
\usepackage{subfig}
\usepackage{textcomp}
\usepackage{balance}
\usepackage{url}

\usepackage{pgfplots}
\usepackage{grffile}
\pgfplotsset{compat=newest}
\usetikzlibrary{plotmarks}
\usetikzlibrary{arrows.meta}
\usepgfplotslibrary{patchplots}

\usepackage[font=small]{caption}

\usepackage{booktabs,array}
\usepackage{mathtools}
\usepackage{bmpsize}
\usepackage{adjustbox}
\usepackage[abs]{overpic}
\usepackage{multirow}
\usepackage{algorithm}
\usepackage{algpseudocode}
\makeatletter
\def\BState{\State\hskip-\ALG@thistlm}
\makeatother

\usepackage{colortbl}%
\newcommand{\myrowcolour}{\rowcolor[gray]{0.925}}

\newcommand{\midsepremove}{\aboverulesep = 0mm \belowrulesep = 0mm}
\midsepremove
\newcommand{\midsepdefault}{\aboverulesep = 0.605mm \belowrulesep = 0.984mm}
\midsepdefault

\newcommand\figHeight{2.5in}
\newcommand\figWidth{4.521in}

\definecolor{mycolor1}{rgb}{0.00000,0.44700,0.74100}%
\definecolor{mycolor3}{rgb}{0.92900,0.69400,0.12500}%
\definecolor{mycolor4}{rgb}{0.46667,0.67451,0.18824}%
\definecolor{mycolor2}{rgb}{0.85000,0.32500,0.09800}%

\definecolor{mycolor0}{rgb}{1,1,1}%
\definecolor{mycolor1}{rgb}{0.00000,0.45000,0.74000}%
\definecolor{mycolor2}{rgb}{0.85000,0.30000,0.10000}%
\definecolor{mycolor3}{rgb}{0.04706,0.47451,0.58039}%
\definecolor{mycolor4}{rgb}{0.00000,0.44700,0.74100}%
\definecolor{mycolor5}{rgb}{0.85000,0.32500,0.09800}%
\definecolor{mycolor6}{rgb}{0.00000,0.44700,0.74100}%
\definecolor{mycolor7}{rgb}{0.85000,0.32500,0.09800}%
\definecolor{mycolor8}{rgb}{0.92900,0.69400,0.12500}%
\definecolor{mycolor9}{rgb}{0.93000,0.69000,0.13000}%
\definecolor{mycolor10}{rgb}{0.49000,0.18000,0.56000}%
\definecolor{mycolor11}{rgb}{0.32000,0.49000,0.08000}%
\definecolor{mycolor12}{rgb}{0.85000,0.33000,0.10000}%
\definecolor{mycolor13}{rgb}{0.30000,0.42000,0.13000}%
\definecolor{mycolor14}{rgb}{0.64000,0.08000,0.18000}%
\definecolor{mycolor15}{rgb}{0.46667,0.67451,0.18824}%
\definecolor{mycolor16}{rgb}{0.49412,0.18431,0.55686}%
\definecolor{mycolor17}{rgb}{0.30100,0.74500,0.93300}%
\newcommand\markSize{2.50000pt}

\newcommand{\myarrowdotted}[1][0.1pt]
{   \begin{tikzpicture}[overlay]
	\draw [->,>=stealth,line width=0.4mm,dashed,black] (-0.1, 0.1) -- (0.4, 0.1);
	\end{tikzpicture}
}

\newcommand{\myarrowsolid}[1][0.1pt]
{   \begin{tikzpicture}[overlay]
	\draw [->,>=stealth,line width=0.4mm,solid,black] (-0.1, 0.1) -- (0.4, 0.1);
	\end{tikzpicture}
}

\definecolor{blueelipse}{rgb}{0,0.2745,1}%
\definecolor{redelipse}{rgb}{0.784,0,0.1254}%
\definecolor{purpleelipse}{rgb}{0.5921,0.0470,0.8745}%
\newcommand{\myelipseblue}[1][0.1pt]
{   \begin{tikzpicture}[overlay]
	\draw [line width=0.5mm,blueelipse](0, 0.1) ellipse (1mm and 1mm);
	\end{tikzpicture}
}
\newcommand{\myelipsered}[1][0.1pt]
{   \begin{tikzpicture}[overlay]
	\draw [line width=0.5mm,redelipse](0, 0.1) ellipse (1mm and 1mm);
	\end{tikzpicture}
}
\newcommand{\myelipsepurple}[1][0.1pt]
{   \begin{tikzpicture}[overlay]
	\draw [line width=0.5mm,purpleelipse](0, 0.1) ellipse (1mm and 1mm);
	\end{tikzpicture}
}

\def\BibTeX{{\rm B\kern-.05em{\sc i\kern-.025em b}\kern-.08em
		T\kern-.1667em\lower.7ex\hbox{E}\kern-.125emX}}
\begin{document}
	\title{Age-Net: An MRI-Based Iterative Framework\\ for Brain Biological Age Estimation}
	\author{Karim Armanious, Sherif Abdulatif, Wenbin Shi, Shashank Salian, Thomas K\"ustner,\\ Daniel Weiskopf, Tobias Hepp, Sergios Gatidis, Bin Yang \IEEEmembership{Senior Member, IEEE}
		\thanks{K. Armanious, S. Abdulatif, W. Shi and B. Yang are with the Institute of Signal Processing and System Theory, University of Stuttgart, 70569 Stuttgart, Germany (e-mail: karim.armanious@iss.uni-stuttgart.de)}
		\thanks{S. Salian and D. Weiskopf are with the Visualization Research Center, University of Stuttgart, 70569 Stuttgart, Germany}
		\thanks{T. K\"ustner and S. Gatidis  are with the Department of Diagnostic and Interventional Radiology, University Hospital Tübingen, 72076 Tübingen, Germany}
		\thanks{T. Hepp is with the Empirical Inference Department, Max Planck Institute for Intelligent Systems, 72076 Tübingen, Germany}
		\thanks{This paper was accepted in part at the IEEE European Signal Processing Conference (EUSIPCO), 2020 \cite{Extra}. }
		\thanks{The first two authors equally contributed to this work.}}
	\maketitle
	
	\begin{abstract}
		The concept of biological age (BA) - although important in clinical practice - is hard to grasp mainly due to the lack of a clearly defined reference standard. For specific applications, especially in pediatrics, medical image data are used for BA estimation in a routine clinical context. Beyond this young age group, BA estimation is mostly restricted to whole-body assessment using non-imaging indicators such as blood biomarkers, genetic and cellular data. However, various organ systems may exhibit different aging characteristics due to lifestyle and genetic factors. Thus, a whole-body assessment of the BA does not reflect the deviations of aging behavior between organs. To this end, we propose a new imaging-based framework for organ-specific BA estimation. In this initial study we focus mainly on brain MRI. As a first step, we introduce a chronological age (CA) estimation framework using deep convolutional neural networks (Age-Net). We quantitatively assess the performance of this framework in comparison to existing state-of-the-art CA estimation approaches. Furthermore, we expand upon Age-Net with a novel iterative data-cleaning algorithm to segregate atypical-aging patients (BA $\not \approx$ CA) from the given population. We hypothesize that the remaining population should approximate the true BA behavior. We apply the proposed methodology on a brain magnetic resonance image (MRI) dataset containing healthy individuals as well as Alzheimer's patients with different dementia ratings. We demonstrate the correlation between the predicted BAs and the expected cognitive deterioration in Alzheimer’s patients. A statistical and visualization-based analysis has provided evidence regarding the potential and current challenges of the proposed methodology.
	\end{abstract}
	
	\vspace{-1mm}
	\section{Introduction}
	\label{sec:introduction}
	\IEEEPARstart{A}{ge} is one of the most important parameters describing individuals in a medical context. For instance, age has a significant impact on the establishment of working diagnoses and the choice of appropriate diagnostic tests \cite{RN1}. Similarly, age is also an important parameter influencing therapeutic decisions in a wide range of clinical situations \cite{RN2,RN3}.
	
	However, age-related biological phenotypes can deviate significantly between individuals within the same age group. These observations have motivated the concept of biological age (BA) in contrast to chronological age (CA) \cite{RN6}. CA is described as the amount of time since the birth of an individual. Unlike CA, BA is not clearly defined. It can be described as a measure for the extent of genetic, metabolic and functional changes in an individual that occur during the process of aging. Thus, BA can be considered as an extension to the traditional concept of CA in addition to any organ-specific accelerated or delayed aging characteristics \cite{RN7,RN8,RN10}. Despite this relatively imprecise definition, the potential impact of the concept of BA on patient management is easily conceivable. It is a common practice for clinicians to assess the overall condition of patients as part of the clinical examination relative to their respective age group and incorporate this impression into their medical decisions. However, these personal estimates are subjective and not easily quantifiable.
	
	As an expansion to the concept of BA, the notion of organ-specific BA has been proposed aiming to describe changes in morphology, biology and function of organ systems that occur with aging \cite{RN11}. This concept is based on the rationale that single organs or organ systems can be affected by different genetic or environmental factors, and thus display different courses of aging. As an example, parameters of pulmonary function were proposed as measures for lung BA \cite{RN13}.
	
	\begin{figure*}[!t]
		\subfloat[Hybrid 3D network model with a stem network followed by a basic building block of two inception blocks and a fire module.\label{fig:whole}]{\hspace{5mm}\includegraphics[width=1.9\columnwidth]{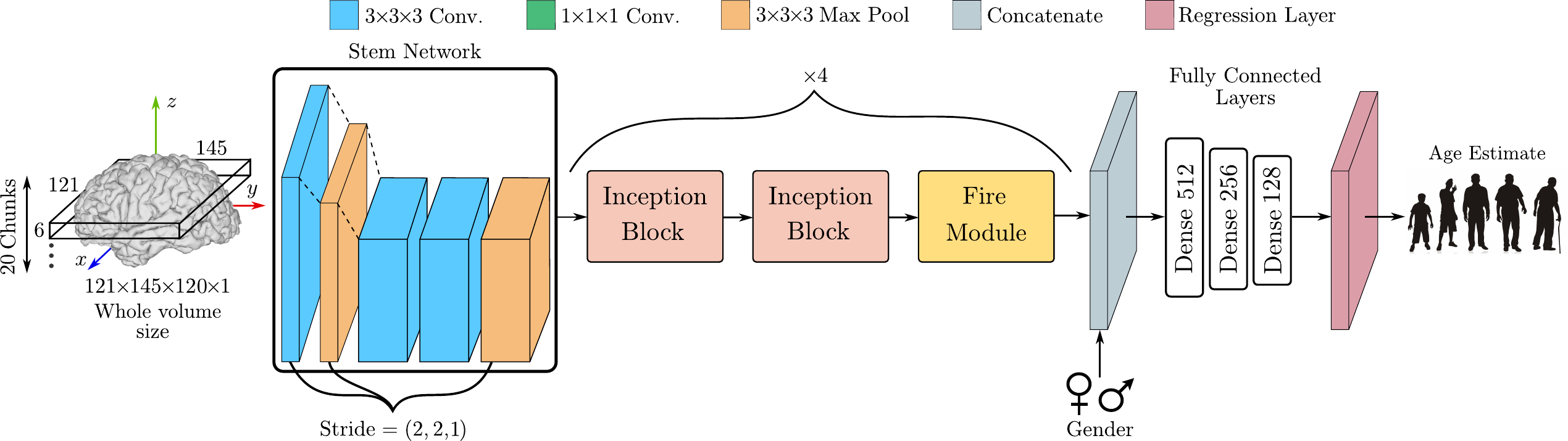}}
		\vspace{1mm}
		\centering
		\subfloat[Inception block basic architecture.]{{\includegraphics[width=0.9\columnwidth]{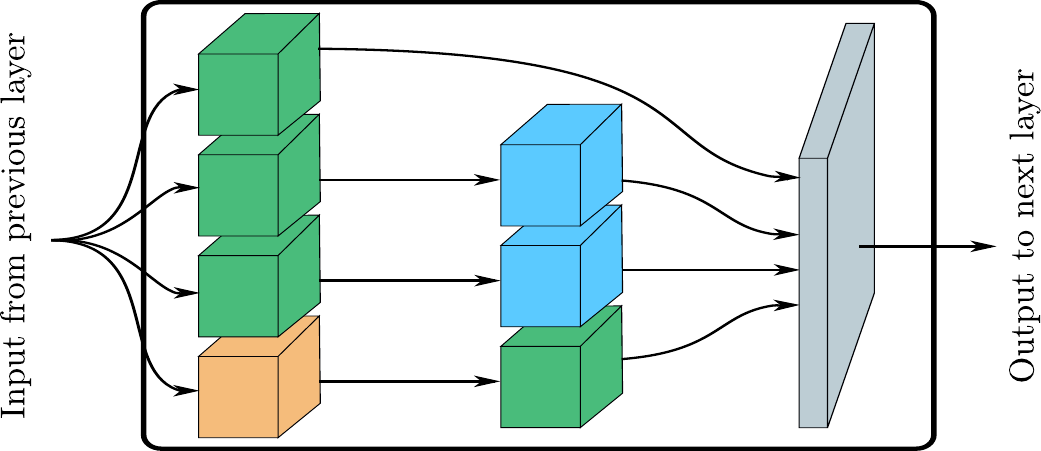} } \label{fig:inception}}%
		\qquad
		\subfloat[Fire module basic architecture.]{{\includegraphics[width=0.9\columnwidth]{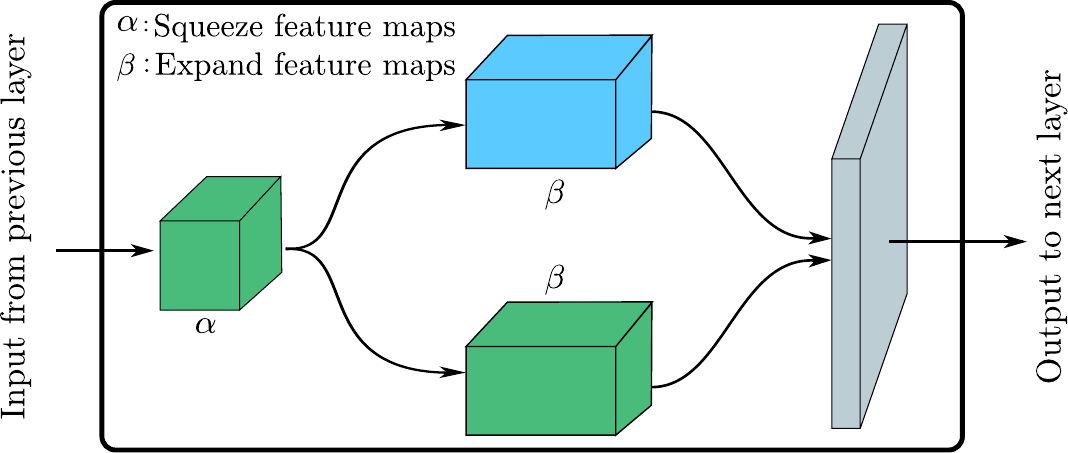} }\label{fig:fire}}%
		\caption{\small An overview of the proposed chronological age estimation network (Age-Net) with the detailed architecture of each block. The network takes as an input either the whole organ volume or volume chunks of segmented brain gray matter (GM). The dimensions of the separate blocks are the same regardless of the utilized data-feeding strategy (full/chunks).\label{fig:architecture}}
		\vspace{-2mm}
	\end{figure*}
	
	A large body of research attempts the quantification of BA using non-imaging data. More specifically, age-dependent variables such as genetic \cite{RN6,RN7}, cellular \cite{RN8}, phenotypic \cite{RN10,RN12} and epidemiological data \cite{RN9}, blood biomarkers \cite{E1,E2} and physical activity \cite{E3} have been used as indicators for the BA. Traditional approaches rely on classical techniques such as multiple linear regression (MLR) \cite{E4} or the Klemera and Doubal (KD) method \cite{E5}, whereas recent works adopt deep neural networks \cite{E6,E7}. The majority of these approaches utilize large cohort datasets for the prediction of the mortality risk \cite{E9,E10}. Other methods incorporate the CA as ground-truth labels and examine the relation between the predicted ages and other health indicators, such as the work ability index (WAI) \cite{E11} and frailty index (FI) \cite{E12}, for assessing the BA.  
	
	Nonetheless, the majority of the above non-imaging approaches lead to a whole-body assessment of the BA. In this sense, they are not capable of recognizing the differences in aging characteristics between individual organ systems. To counteract this limitation, several current approaches rely on the utilization of multiple non-imaging biomarkers, such as liver and lung function data, to provide a more specific BA estimation for multiple organs conjointly \cite{E9}. With this perspective, medical imaging data may also provide significant information allowing for estimation of organ-specific BA.

	The use of medical imaging for age estimation in a clinical setting is limited mainly to skeletal age estimation in infants and adolescents using conventional radiography and MRI \cite{RN14,Final1,x18-2,x18-3}.
	Beyond skeletal age estimation, first studies have introduced the concept of brain age based on changes in brain morphology (e.g., changes in subvolumes) and provided evidence for associations between premature brain aging and cognitive function \cite{RN15,Final2}. 
	
	Brain age can be defined by comparing an individual’s morphological brain features to a reference database of the underlying population. This is possible using brain MRI due to the highly constant anatomy of the central nervous system that allows for the detection of age-dependent morphological deviations. For example, J.~H.~Cole was the first to demonstrate the correlation between brain age and the risk of mortality using a combination of MRI data and DNA-methylation \cite{E8}. Similarly, several disorders were found to associate with abnormal brain aging. For instance, the authors in \cite{Resub-1,Resub-2,Resub-3} have demonstrated the impact of traumatic brain injury, refractory focal epilepsy and schizophrenia on accelerated brain aging. Moreover, increased apparent brain aging was observed in HIV-positive patients \cite{Resub-4}. Similarly, Alzheimer’s disease was established to correlate with abnormal brain aging \cite{E16}.
	
	More recent efforts have incorporated the use of machine learning (ML) and deep learning (DL) techniques for medical imaging-based age estimation. Traditional approaches, such as kernel methods \cite{x20} and support vector machines (SVMs) \cite{x21}, were initially utilized for brain CA estimation using T1-weighted MRI volumes. Also, atlas-based methods were employed to extract effective local features for the same task \cite{x22}. However, in recent years, the use of convolutional neural networks (CNNs) has become more prevalent due to strong results in a multitude of medical tasks, such as classification and segmentation \cite{x24,Extra2,x25,Extra8}. For instance, a deep CNN has been utilized for prediction of brain age using 2D T1-weighed MR images \cite{x11}. This framework was then expanded upon to incorporate manually extracted features in addition to the 2D CNN architecture \cite{x26}. Recent advances have attempted the use of relatively shallow 3D CNN architectures to incorporate the spatial information between slices in the brain age estimation procedure \cite{x27,x15}. In 2020 \cite{Resub-5}, authors proposed a deeper 3D architecture consisting of 7 convolutional layers leading to state-of-the-art results in MRI-based brain age estimation. For forensic applications, a large body of research has utilized hand and skeletal MRI volumes to estimate the BA \cite{x16,x17,x18}.
	
	As stated clearly in the most recent survey on this topic \cite{E13}, all DL approaches, whether imaging or non-imaging based, relies on the CA as ground-truth labels for BA prediction \cite{E14,x27}. Thus, the predicted ages cannot be used to assess the true aging characteristics of the test subjects. To the best of our knowledge, the problem of defining BA ground-truth labels is still not possible and remains an open research question.
	
	The purpose of this study is to bridge the gap between chronological and biological age estimation. This is achieved by, first, introducing a new DL framework for brain CA estimation using 3D medical imaging data. The performance of this framework is validated by a quantitative comparison with other state-of-the-art CA estimation networks for brain age estimation. Additionally, a novel iterative training strategy is presented as an initial solution for approximating brain BA labels. This is achieved by identifying outliers who exhibit atypical-aging characteristics. These outliers are then segregated from the training dataset. A serious challenge presents itself on how to validate the accuracy of the utilized training approach. To this end, we apply the proposed methodology on a dataset containing both healthy and Alzheimer's patients. Subsequently, we quantify the amount of Alzheimer's patients detected as atypically-aging by the iterative strategy. Statistical and visual analysis of the results is conducted to illustrate the merit and limitations of the proposed methodology.
	
	This paper is organized as follows: Sec.~\ref{sec:ca} describes the proposed CA estimation network together with the conducted comparative study and the corresponding results. Sec.~\ref{sec:ba} presents the iterative data-cleaning strategy for BA estimation and describes the conducted experimental evaluations. Finally, Sec.~\ref{sec:results} presents the results and discussions for the BA framework followed by the conclusion in Sec.~\ref{sec:conc}.
	
	\section{Chronological Age Estimation}\label{sec:ca}
	In this section, the proposed DL architecture for organ-specific CA estimation is introduced. This network is later utilized as the foundation of the proposed iterative training strategy for BA estimation. A detailed description of the experiments conducted to validate the proposed architecture is presented.

	\begin{table}[!t]
		\caption{\small \\Age-Net Architecture. \label{tab:architecture}}
		\centering
		\setlength\arrayrulewidth{0.05pt}
		\small
		\bgroup
		\midsepremove
		\def\arraystretch{1.15}		
		\resizebox{\columnwidth}{!}{%
			\begin{tabular}{l cc}
				\toprule
				Layer & Output shape & \# Parameters\\ 
				\midrule  
				\myrowcolour Input (Brain 3D chunk) & $121 \times 145 \times 6 \times 1$ & ---\\\midrule
				
				Inception stem network & $16 \times 19 \times 6 \times 192$ & 445k\\\midrule
				
				\myrowcolour Inception block - 3b & $16 \times 19 \times 6 \times 256$ & 1.42M\\
				\myrowcolour Inception block - 3c & $16 \times 19 \times 6 \times 480$ & 3.18M\\
				\myrowcolour Fire Module ($\alpha$ = 16, $\beta$ = 64) & $16 \times 19 \times 6 \times 128$ & 36.8k\\\midrule
				
				Inception block - 4b & $8 \times 10 \times 6 \times 512$ & 5.33M\\
				Inception block - 4c & $8 \times 10 \times 6 \times 512$ & 1.96M\\
				Fire Module ($\alpha$ = 16, $\beta$ = 64) & $8 \times 10 \times 6 \times 128$ & 36.8k\\\midrule
				
				\myrowcolour Inception block - 4d & $8 \times 10 \times 6 \times 512$ & 5.68M\\
				\myrowcolour Inception block - 4e & $8 \times 10 \times 6 \times 528$ & 6.04M\\
				\myrowcolour Fire Module ($\alpha$ = 16, $\beta$ = 64) & $8 \times 10 \times 6 \times 256$ & 141k\\\midrule
				
				Inception block - 4f & $4 \times 5 \times 3 \times 832$ & 9.70M\\
				Inception block - 5b & $4 \times 5 \times 3 \times 832$ & 5.47M\\
				Fire Module ($\alpha$ = 32, $\beta$ = 256) & $4 \times 5 \times 3 \times 512$ & 524k\\\midrule
				
				\myrowcolour Global average pooling & 512 & ---\\\midrule
				
				Dense layer 1 & 512 & 260k\\
				Dense layer 2 & 256 & 130k\\
				Dense layer 3 & 128 & 32k\\\midrule
				
				\myrowcolour Regression layer & 1 & 129\\\midrule
				
				\# parameters (total) & --- & 41M\\
				\bottomrule
			\end{tabular}
		}
		\egroup
		\vspace{-2mm}
	\end{table}

	\subsection{Architectural Details}
	The proposed network for CA estimation, hereby referred to as Age-Net, is illustrated in Fig.~\ref{fig:whole}. Based on an investigative assessment of different state-of-the-art DL structures, including ResNeXt \cite{x1} and DenseNets \cite{x2} among others, the proposed regression network was constructed out of a hybrid combination of inception v1 \cite{x3} and SqueezeNet \cite{x4} architectures. 
	
	The inception modules are based on the concept of split-transform-merge strategy where each module is comprised of parallel filters with different kernel dimensionality, which results in a network growing wider instead of deeper. This enables learning deeper feature representations by increasing the capacity of the network while mitigating the increased computation budget associated with network depth \cite{x3}. In this work, we utilize the inflated inception v1 architecture, which is a 3D realization of conventional inception modules achieved via inflating all filters and pooling kernels into their 3D counterparts \cite{x5}. An illustration of an inflated inception module is illustrated in Fig.~\ref{fig:inception}.

	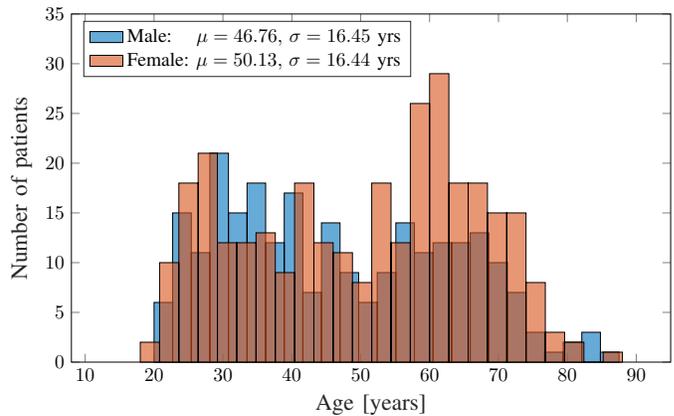
\begin{figure}
		\centering
		\resizebox{\columnwidth}{!}{
%
%
\definecolor{mycolor1}{rgb}{0.00000,0.44700,0.74100}%
\definecolor{mycolor2}{rgb}{0.85000,0.32500,0.09800}%
\begin{tikzpicture}
\begin{axis}[%
width=4.3in,
height=\figHeight,
at={(0.758in,0.481in)},
scale only axis,
xmin=8,
xmax=95,
xlabel style={font=\color{white!15!black}},
xlabel={\large Age [years]},
ymin=0,
ymax=35,
ylabel style={font=\color{white!15!black}},
ylabel={\large Number of patients},
axis background/.style={fill=white},
title style={font=\color{white!15!black},yshift=-6},
legend style={at={(0.02,0.98)}, anchor=north west, legend cell align=left, align=left, draw=white!15!black}
]
\addplot[ybar interval, fill=mycolor1, fill opacity=0.6, draw=black, area legend] table[row sep=crcr] {%
	x	y\\
	20	6\\
	22.7	15\\
	25.4	11\\
	28.1	21\\
	30.8	15\\
	33.5	18\\
	36.2	12\\
	38.9	17\\
	41.6	7\\
	44.3	14\\
	47	9\\
	49.7	6\\
	52.4	9\\
	55.1	14\\
	57.8	11\\
	60.5	12\\
	63.2	12\\
	65.9	13\\
	68.6	10\\
	71.3	7\\
	74	3\\
	76.7	1\\
	79.4	2\\
	82.1	3\\
	84.8	1\\
	87.5	1\\
};
\addlegendentry{$\text{Male:  \:\,\, }\mu = 46.76, \, \sigma = 16.45 \text{  yrs}$}

\addplot[ybar interval, fill=mycolor2, fill opacity=0.6, draw=black, area legend] table[row sep=crcr] {%
	x	y\\
	18	2\\
	20.8	10\\
	23.6	18\\
	26.4	21\\
	29.2	12\\
	32	12\\
	34.8	13\\
	37.6	9\\
	40.4	18\\
	43.2	12\\
	46	11\\
	48.8	8\\
	51.6	18\\
	54.4	12\\
	57.2	26\\
	60	29\\
	62.8	18\\
	65.6	18\\
	68.4	15\\
	71.2	15\\
	74	8\\
	76.8	3\\
	79.6	2\\
	82.4	0\\
	85.2	1\\
	88	1\\
};
\addlegendentry{$\text{Female: }\mu = 50.13, \, \sigma = 16.44 \text{  yrs}$}

\end{axis}
%
\end{tikzpicture}%
}
		\caption{\small Distribution of the open-source brain IXI MR dataset. The dataset covers diverse age categories with a mean of $\mu \approx 48$ years and a standard deviation of $\sigma \approx 16$ years. \label{fig:ageDist}}
		\vspace{-4mm}
	\end{figure}

	Additionally, fire modules proposed in SqueezeNet are also incorporated in the Age-Net architecture \cite{x4}. They consist of squeeze-and-expand layers comprised of a combination of $1 \hspace{0.08mm} \times \hspace{0.08mm} 1 \hspace{0.08mm} \times \hspace{0.08mm} 1$ and $3 \times 3 \times 3$ convolutions which helps to reduce the total number of trainable parameters while enhancing the representation capacity. An illustration of the fire modules is depicted in Fig.~\ref{fig:fire}.
	
	The final architecture for Age-Net is composed of an initial stem network consisting of $3 \hspace{0.4mm} \times \hspace{0.4mm} 3 \hspace{0.4mm} \times \hspace{0.4mm} 3$ convolutions and max-pooling layers. This is followed by four modules concatenated together in an end-to-end manner. Each module consists of two inception blocks followed by a single fire module. In the above modules, each convolutional layer is followed by batch normalization and a ReLU activation function. L2 regularization is additionally utilized for each convolutional layer. A global average pooling layer is then applied as a structural regularizer to reduce the four-dimensional tensor to a one-dimensional output vector of 512 features. Finally, three dense layers combine additional gender labels (male/female) before a final regression layer outputs the predicted age. This is based on prior work that discussed differences between male and female brain aging \cite{Resub-6}. Thus, we chose to include the gender labels into our proposed architecture. The complete architectural details of the Age-Net are outlined in Table~\ref{tab:architecture}. Kernel parameters for the different inception blocks can be found in the original inception v1 publication \cite{x3}.
	\subsection{Input Pipeline}
	
	Empirically, the method of feeding the input MR volumes to Age-Net was found to have a significant impact on the network performance. Thus, two different approaches of feeding the 3D volumes are investigated in this work. First, entire MR volumes are fed as inputs to the network, which produces a single predicted age for each test subject. Due to the relatively limited number of training patients, data augmentation is essential to prevent network overfitting. Accordingly, horizontal flipping and translating the input volumes within a pre-defined voxel-range ($30 \times 30$) in the axial orientation were incorporated in the input pipeline. Despite its simplicity, this approach comes with a significant cost on the training efficiency due to the large memory space required as well as the on-the-fly data augmentations.
	
	The second data feeding strategy entails dividing the input MR volumes into smaller 3D chucks and then feeding each chunk separately. This implicitly augments the training process by expanding the number of input samples to the network, thus, negating the need for on-the-fly data augmentations. The final predicted score for a test subject is then given as the mean of the predicted ages for the different input chunks. In addition to being advantageous from a training efficiency perspective, we hypothesize that this approach could assist in the unsupervised localization of anomalies and lesions per chunk. This can be achieved by investigating irregularities in the prediction scores for individual chunks compared to the CA ground-truth label for each test subject. This hypothesis will be further investigated in the future.
	
	\begin{figure}
		\centering
		\subfloat[]{\includegraphics[width = 0.4\columnwidth,height = 0.35\columnwidth]{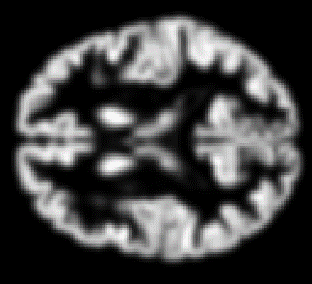}}
		\hspace{6mm}
		\subfloat[]{\includegraphics[width = 0.4\columnwidth,height = 0.35\columnwidth]{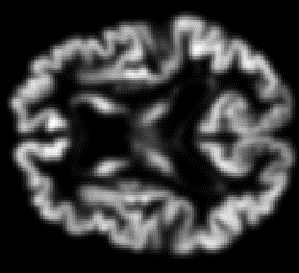}}
		\vspace{-1mm}\caption{\small Examples of input brain GM from the axial orientation: \\(a)~23 year-old subject, (b)~86 year-old subject. \label{fig:examri}}
		\vspace{-2.25mm}
	\end{figure}
	
	\subsection{Dataset and Pre-processing}\label{sec:data-ca}
	
	The proposed Age-Net architecture is evaluated on the task of estimating organ-specific CA for the brain region. For this purpose, we utilize the open-source IXI dataset for brain MR scans \cite{x6}. T1-weighted MR scans with an original matrix size of $256 \times 256 \times 150$ voxels from 562 healthy subjects with no neurological disorders were utilized. The scans were acquired according to a flip angle of 8°, echo time of 4.6 ms and a repetition time of 9.8 ms. Scans from 420, 92 and 50 subjects were used for training, validation and testing, respectively. This was achieved while maintaining a balanced number of scans from all age-groups in all data subsets. For pre-processing, we utilize the steps recommended in \cite{x8,Extra9}. As a first step, all 3D volumes were realigned to provide a common orientation for brain visualization. This was achieved by utilizing the canonical Montreal Neurological Institute (MNI) 152 template adopted by the international consortium for brain mapping (ICBM) \cite{x8}. Since the gray matter (GM) content of the brain was previously proven to be a strong indicator of the brain age \cite{x19}, the GM content of the aligned brain volumes were segmented using the statistical parameter mapping 12 (SPM12) software \cite{x9}. The resulting tissue maps were registered using DARTEL \cite{x10} followed by normalization and modulation using a Jacobian deformation map. The resultant final GM volumes are of size $121 \times 145 \times 121$ voxels. All the following experiments were conducted using these GM volumes. The CA histogram for the brain MR volumes are depicted in Fig.~\ref{fig:ageDist} and example images are illustrated in Fig.~\ref{fig:examri}. 
	
	\vspace{-1.5mm}\subsection{Experiments}
	
	The proposed Age-Net architecture for CA estimation is investigated for the two different data feeding strategies described above. The first variant (Age-Net-Volume) involves feeding the entire MR volume as input to the network. In contrast, the second variant (Age-Net-Chunk) feeds smaller 3D chunks as inputs. Each volume is divided into 20 successive chunks of matrix size $121 \times 145 \times 6$. However, the first and last 3 chunks (spanning the first and last 18 slices) were found not to contain significant GM content. This is because they mainly represent either the top or base of the skull region with little GM information. As such, we have chosen to consider the middle 14 chunks in our age estimation procedure as they contain the vast majority of important brain regions. Additionally, we hypothesize that including additional meta-information about the test subjects, i.e., gender labels, would assist in enhancing the age-regression performance \cite{Resub-6}. As such, an additional experiment was conducted to investigate the effect of including the gender (Age-Net-Gender) with the chunk data feeding strategy.

		\begin{table}[!]
		\caption{\small \\Quantitative comparison for CA estimation on IXI dataset.\label{tab:results}}
		\centering
		\setlength\arrayrulewidth{0.05pt}
		\small
		\bgroup
		\midsepremove
		\def\arraystretch{1.45}
		\resizebox{\columnwidth}{!}{%
			\begin{tabular}{l ccccc}
				\toprule
				Model & MAE & SD & Bias & RMSE & Corr.\\ 
				\midrule  
				\myrowcolour 2D-Huang \cite{x11} & 3.529 & 4.302 & 1.250 & 4.480 & 0.969\\
				3D-Ueda \cite{x12} & 3.705 & 4.298 & 1.268 & 4.481 & 0.969\\
				\myrowcolour 3D-Peng \cite{Resub-5} & 2.741 & 3.690 & 0.517 & 3.726 & 0.981\\
				Age-Net-Volume & 2.658 & 3.532 & 0.608 & 3.584 & 0.979 \\
				\myrowcolour Age-Net-Chunk & 2.283 & 3.546 & 0.902 & 3.659 & 0.978 \\
				Age-Net-Gender & \textbf{1.955} & \textbf{3.169} & \textbf{0.511} & \textbf{3.210} & \textbf{0.983} \\
				\bottomrule
			\end{tabular}
		}
		\egroup
		\vspace{-2.5mm}
	\end{table}
	
	\begin{figure*}
		\centering
		\resizebox{\textwidth}{!}{\input{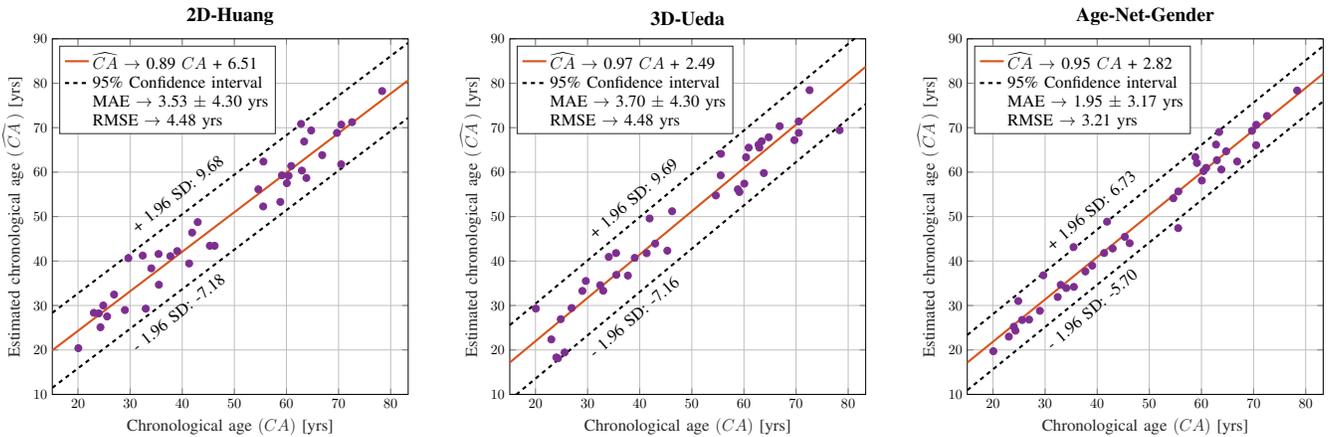}}
		\caption{\small Statistical analysis and comparison between some networks proposed in literature and the proposed Age-Net with the chunk-based feeding strategy. The Age-Net approach outperforms the concurrent MRI-based CA estimation approaches across the  investigated metrics. \label{fig:resPlots}}
		\vspace{-1mm}
	\end{figure*}

	\begin{figure}
		\centering
		\includegraphics[width=1\columnwidth]{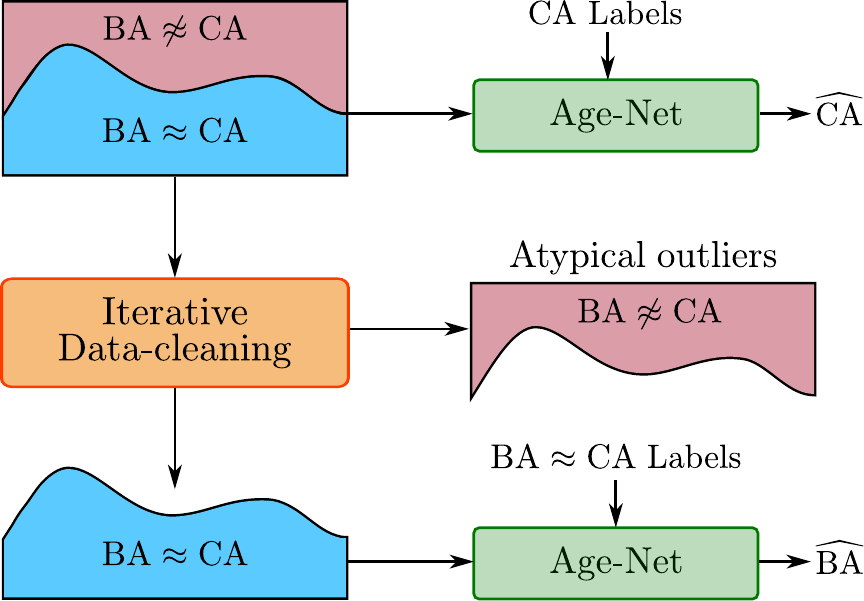}
		\caption{\small An overview of the iterative data-cleaning strategy where atypical outliers ($\textrm{BA} \not \approx \textrm{CA}$) are segregated from the training data. Thus, the Age-Net (with the chunk data feeding strategy) is trained with only typical-aging dataset ($\textrm{BA} \approx \textrm{CA}$).\label{fig:data_cleaning}}
		\vspace{-3.5mm}
	\end{figure}

	To demonstrate the performance of Age-Net, quantitative comparisons were conducted against other regression networks previously proposed for brain MRI CA estimation. First, we compared against the framework provided in \cite{x11} (2D-Huang) that consists of a modified deep VGG-Net \cite{x13}.  Also, comparisons were conducted against a brain age regression framework (3D-Ueda) \cite{x12}. This network is comprised of four 3D convolutional blocks together with max pooling and dense layers. Finally, we examine the performance of Age-Net in comparison to a recent brain age regression network proposed by \cite{Resub-5}. This framework (3D-Peng) utilizes 3D MRI volumes together with a fully convolutional architecture to achieve state-of-the-art brain age regression results. All networks were trained until convergence to minimize the mean absolute error (MAE) loss function on a single NVIDIA Titan-X GPU using the Adam optimizer \cite{x14} with Nesterov momentum (0.9) and a learning rate of $10^{-4}$. All implementations of Age-Net will be made publicly available upon the publication of this work\footnote{https://github.com/KarimArmanious/Age-Net}. Several metrics were calculated for the quantitative comparisons: the MAE, standard deviation (SD), bias, root mean square error (RMSE) and the correlation coefficient (Corr.) between the predicted ages and the ground-truth CA labels.
	
	\vspace{-1.5mm}\subsection{Results}
	
	The results of the CA estimation for the IXI brain dataset are presented in Table~\ref{tab:results} and Fig.~\ref{fig:resPlots}. The current approaches by Huang \cite{x11} and Ueda \cite{x12} exhibit comparable performance with an MAE of approximately 3.5 years. The proposed 3D Age-Net architecture as well as the recent 3D-Peng framework \cite{Resub-5} outperform the prior two approaches across the utilized metrics. They similarly achieve comparable performance with an MAE of around 2.7 years when dealing with full GM volumes. However, adapting the input pipeline of Age-Net to accommodate smaller 3D chunks instead of full volumes improved the MAE approximately by $0.4$ years albeit with an increased bias of $0.9$ years. Furthermore, including the gender labels with the chunk data feeding strategy resulted in the best quantitative scores represented by MAE of $1.96$ years, the lowest systematic error of $0.51$ years in bias and the smallest RMSE of $3.2$ years. An interesting observation regarding the quantitative metrics is that all different CA estimation approaches result in positive bias values. This indicates the tendency of the investigated subjects to exhibit accelerated aging characteristics. Additional results examining the performance of the Age-Net architecture on a different brain dataset (OASIS-3) are presented in Appendix~A.
	
		\begin{figure*}[t]
		\centering
		\includegraphics[width=0.95\textwidth]{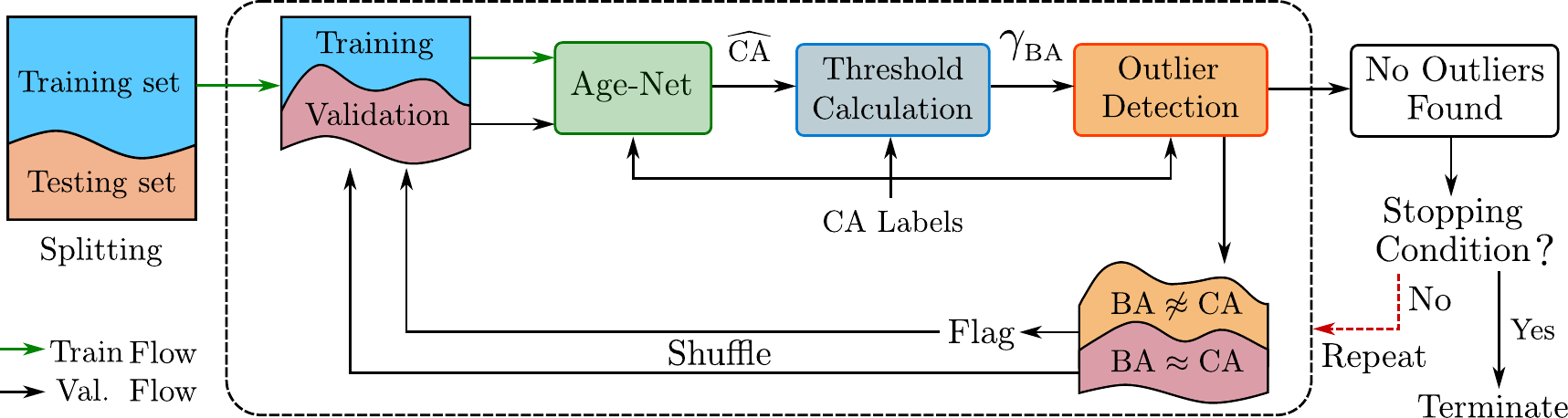}
		\caption{\small A detailed flow chart of the proposed iterative data-cleaning strategy for the extraction of BA labels. \label{fig:iterative_cleaning}}
		\vspace{-2mm}
	\end{figure*} 

	\section{Biological Age Estimation}\label{sec:ba}
	
	The main challenge for image-based BA estimation is the lack of ground-truth labels since BA is not clearly defined \cite{E6,E7}. As such, previous approaches attempting this task had to, instead, rely on utilizing CA labels \cite{x16,x17,x18}. These approaches interpret the CA prediction errors as a sign of abnormal aging. However, this is correct only under the assumption that the investigated training population is healthy and have reliable CA labels. This assumption is no longer valid if a portion of the investigated population suffers from a disease. Also, this does not apply to general population studies where individuals may have undiagnosed disorders. Accordingly, in this work we hypothesize that the absence of a diagnostic disease does not guarantee that the addressed population exhibit typical aging characteristics (BA $\approx$ CA). Also, aging is an organ-specific process affected by a multitude of factors such as lifestyle and genetics. Thus, utilizing CA as ground-truth will not provide results that are indicative of the true aging features of the organs in question. Another option is to rely on subjective evaluations by radiologists. Nevertheless, this time-consuming and subjective process is challenging for large datasets and depends on the relative experience of the radiologists as it is not possible to accurately label the BA.
	
	To resolve this challenge, we propose an iterative data-cleaning strategy to approximate BA labels. This is achieved by iteratively identifying and subsequently removing outliers that exhibit atypical-aging characteristics, whether accelerated $\left(\textrm{BA}  > \textrm{CA}\right)$ or delayed aging $\left(\textrm{BA}  < \textrm{CA}\right)$. The rationale behind this approach is to arrive at a typical-aging dataset in which the CA labels resemble the true BA labels $\left(\textrm{BA} \approx \textrm{CA}\right)$. We hypothesize that training on this dataset should help to bridge the gap from chronological to biological age estimation. Since the Age-Net-Gender framework was proven in the previous section to result in the best quantitative CA estimation, we herby utilize this approach as the baseline for the following BA estimation approach. For simplicity, we refer to this architecture as ``Age-Net''. A basic outline of this strategy is depicted in Fig.~\ref{fig:data_cleaning}. In the next sections, the introduced iterative data-cleaning strategy and the outlier detection procedure are further defined.

	\vspace{-2mm}\subsection{Iterative Data-Cleaning Strategy}
	
	A step-wise overview of the iterative data-cleaning strategy for BA estimation is illustrated in Fig.~\ref{fig:iterative_cleaning}. Initially, the available brain MR scans were divided into training and testing datasets. Care was taken so that the training subjects represent the entirety of the available age spectrum in a balanced manner.
	
	For each iteration, the first step is to shuffle and split the training data into two subsets. The first subset, referred to as the ``training'' subset, is used to train an Age-Net architecture utilizing the chunk data feeding strategy with gender labels (Age-Net-Gender) till convergence. This input pipeline was chosen as it showed the best results for CA estimation, as described above in Sec. II. The trained model is then validated on the second subset (the ``validation'' subset) and the estimated CAs $( \, {\widehat{\textrm{CA}}} \,)$ for the different patients is used to calculate a patient-dependent threshold $\gamma_{\textrm{BA}}$. This threshold is then utilized for the detection of outliers that exhibit atypical-aging characteristics in the validation subset. The process of threshold calculation and outlier detection is explained in more details in the next subsection. The identified patients are then flagged as outliers. A new iteration would then be repeated starting with merging the validation samples with the training subset, reshuffling and repeating the process by training the Age-Net framework from scratch.

	At the end of an iteration, if no outliers are detected in the validation subset, two arguments could be presented. First, the dataset has been thoroughly filtered out with all atypically-aging patients identified as outliers. Thus, no further refinement of the dataset is possible. Another explanation is that despite the lack of outliers in the validation data, some could still exist in the training subset. To protect against this possibility, an empirical stopping condition is enabled that states that three consecutive data-cleaning iterations, trained with different initializations, must yield no new outliers before the iterative strategy can be terminated.
	
	Upon termination of the data-cleaning algorithm, all patients who were flagged as outliers in more than one iteration are removed from the training dataset. This serves to assert that no typically-aging patient is wrongfully detected as an outlier. Also, this process assists in maintaining the training data distribution during the data-cleaning strategy. Finally, an Age-Net architecture is trained on the cleaned dataset (after the removal of the outliers) where the CA labels should correspond approximately to the true BA labels ($\textrm{BA} \approx \textrm{CA}$).
	\begin{figure}[!b]
		\vspace{-5mm}
		\resizebox{0.48\textwidth}{!}{\input{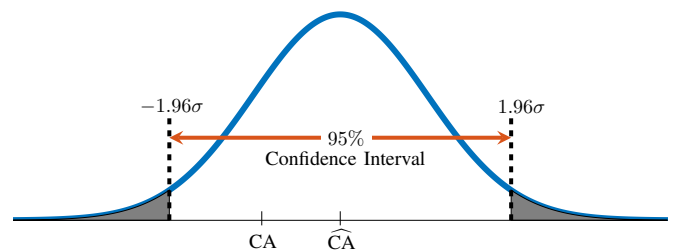}}
		\caption{\small Illustration of the outlier detection threshold $\gamma_{\textrm{BA},n}$. \label{fig:th_illustartion}\vspace{-0.5mm}}
	\end{figure}
	\vspace{-1.5mm}\subsection{Outlier Detection}
	
	In this work, we utilize a chunk data feeding strategy were each input MR volume is divided into $K$ smaller chunks before being fed as input to the Age-Net. Outlier detection is initiated by first calculating a consolidated CA estimate for each patient in the validation dataset. This is achieved by averaging out the predicted ages $(\widehat{\textrm{CA}}_{n,i})$ for each chunk $i$ in the MR volume of patient $n$ as:
	\begin{equation}
	\widehat{\textrm{CA}}_n = \frac{1}{K} \sum_{i=1}^{K} \widehat{\textrm{CA}}_{n,i}
	\end{equation}
	Additionally, the standard deviation for the different chunk predictions is calculated for each patient as:
	\begin{equation}
	\sigma_n= \sqrt{ \frac{1}{K-1} \sum_{i=1}^{K} \left(\widehat{\textrm{CA}}_{n,i} -  \widehat{\textrm{CA}}_n \right)^2}
	\end{equation}
	This is repeated for all patients to obtain the vectors:
	\begin{equation}
	\underline{{\widehat{\textrm{CA}}}} = \begin{pmatrix}
	\widehat{\textrm{CA}}_1 \\
	\widehat{\textrm{CA}}_2 \\
	\vdots  \\
	\widehat{\textrm{CA}}_N 
	\end{pmatrix} \, \, \, \, \, \,  \,,   \,\, \, \, \, \, \, \underline{\mathbf{\sigma}} = \begin{pmatrix}
	\sigma_{1} \\
	\sigma_{2} \\
	\vdots  \\
	\sigma_{N} 
	\end{pmatrix}
	\end{equation}
	where $N$ is the total number of patients in the validation dataset. 
	For outlier detection, we compare the predicated age deviations ($\underline{\textrm{D}}$) against a patient-dependent threshold ($\underline{\gamma_{\textrm{BA}}}$), both defined as:
	\begin{equation}
	\underline{\textrm{D}} = \left| \underline{{\widehat{\textrm{CA}}}} - \underline{{{\textrm{CA}}}}\right| \, \, \, \, \, \,  \,,   \,\, \, \, \, \, \,	\underline{\gamma_{\textrm{BA}}} = R \cdot \underline{\mathbf{\sigma}}
	\label{eq:th}
	\end{equation}
	where $R$ is a pre-defined constant value. The $n^{\textrm{th}}$ patient is flagged as an outlier only if the age deviation exceeds the corresponding threshold value:
	\begin{equation}
	{\textrm{D}}_n > \gamma_{\textrm{BA},n}
	\end{equation}

	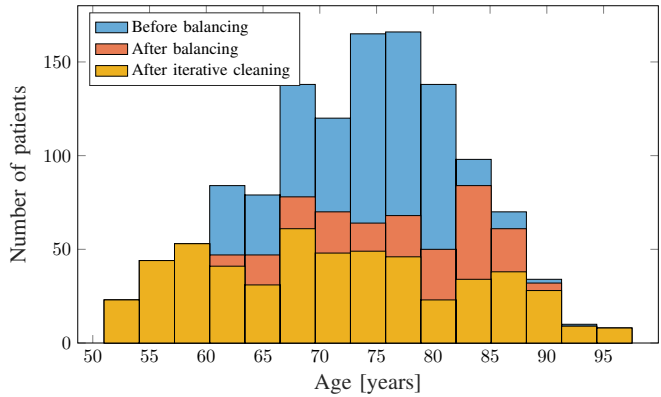
\begin{figure}[!t]
		\resizebox{0.48\textwidth}{!}{
%
%

\definecolor{mycolor1}{rgb}{0.00000,0.44700,0.74100}%
\definecolor{mycolor2}{rgb}{0.91000,0.48500,0.32800}%
\definecolor{mycolor3}{rgb}{0.92900,0.69400,0.12500}%
\begin{tikzpicture}

\begin{axis}[%
width=4.3in,
height=2.5in,
at={(0.758in,0.481in)},
scale only axis,
xmin=48.675,
xmax=99.825,
ymin=0,
ymax=180,
axis background/.style={fill=white},
xlabel style={font=\color{white!15!black}},
xlabel={\large Age [years]},
ylabel style={font=\color{white!15!black}},
ylabel={\large Number of patients},
legend style={at={(0.02,0.98)}, anchor=north west, legend cell align=left, align=left, draw=white!15!black}
]
\addplot[ybar interval, fill=mycolor1, fill opacity=0.6, draw=black, area legend] table[row sep=crcr] {%
x	y\\
51	23\\
54.1	44\\
57.2	53\\
60.3	84\\
63.4	79\\
66.5	138\\
69.6	120\\
72.7	165\\
75.8	166\\
78.9	138\\
82	98\\
85.1	70\\
88.2	34\\
91.3	10\\
94.4	8\\
97.5	8\\
};
\addlegendentry{\small Before balancing}

\addplot[ybar interval, fill=mycolor2, fill opacity=1, draw=black, area legend] table[row sep=crcr] {%
x	y\\
51	23\\
54.1	44\\
57.2	53\\
60.3	47\\
63.4	47\\
66.5	78\\
69.6	70\\
72.7	64\\
75.8	68\\
78.9	50\\
82	84\\
85.1	61\\
88.2	32\\
91.3	9\\
94.4	8\\
97.5	8\\
};
\addlegendentry{\small After balancing}
\label{fig:red}

\addplot[ybar interval, fill=mycolor3, fill opacity=1, draw=black, area legend] table[row sep=crcr] {%
	x	y\\
51	23\\
54.1	44\\
57.2	53\\
60.3	41\\
63.4	31\\
66.5	61\\
69.6	48\\
72.7	49\\
75.8	46\\
78.9	23\\
82	34\\
85.1	38\\
88.2	28\\
91.3	9\\
94.4	8\\
97.5	8\\
};
\addlegendentry{\small After iterative cleaning}

\end{axis}
\end{tikzpicture}%
}
		\caption{\small Histogram of the utilized subset from the OASIS-3 MR dataset. Data balancing was performed with respect to the number of data samples in the different age groups to ensure the Age-Net is trained on an age-balanced dataset. The remaining data samples were allocated to the test dataset. After the iterative cleaning strategy, the remaining population for training the final model is also depicted. \label{fig:oasis_dist}}
		\vspace{-3.3mm}
	\end{figure}
	
	Assuming a normal distribution for the chunk predictions, the constant $R$ was set to $1.96$ to reflect the $95$\% confidence interval of the mean predicted age of each patient, as illustrated in Fig.~\ref{fig:th_illustartion}. This $R$ was chosen based on the experiments presented in Appendix~B. At the end of each iteration, the training and validation datasets are reshuffled and a new iteration would commence until the stopping condition is reached. Upon termination of the iterative data-cleaning, all patients detected as outliers would be removed from the final training dataset only if they were flagged in more than one iteration. The final framework is then trained on a dataset containing only patients exhibiting typical-aging characteristics.
	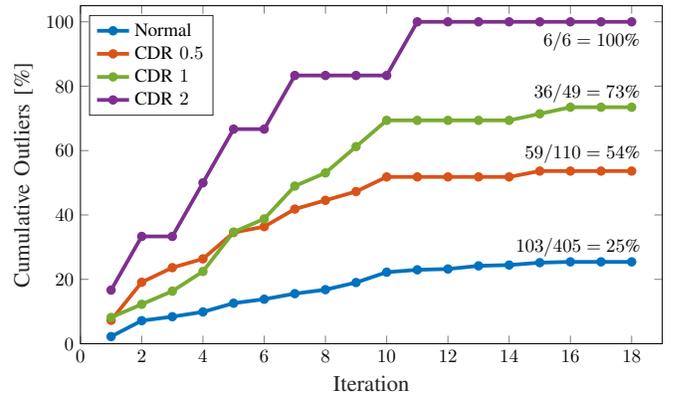
\begin{figure}[!t]
		\resizebox{0.48\textwidth}{!}{
%
%
%
\begin{tikzpicture}

\begin{axis}[%
width=4.3in,
height=2.5in,
at={(0in,0in)},
scale only axis,
xmin=0,
xmax=19,
xlabel style={font=\color{white!15!black}},
xlabel={\large Iteration},
ymin=0,
ymax=105,
ylabel style={font=\color{white!15!black}},
ylabel={\large Cumulative Outliers [\%]},
axis background/.style={fill=white},
legend style={at={(0.015,0.675)}, anchor=south west, legend cell align=left, align=left, draw=white!15!black}
]
\addplot [color=mycolor4, line width=2.0pt, mark size=1.5pt, mark=*, mark options={solid, fill=mycolor4, mycolor4}]
  table[row sep=crcr]{%
1	2.22222222222222\\
2	7.16049382716049\\
3	8.39506172839506\\
4	9.87654320987654\\
5	12.5925925925926\\
6	13.8271604938272\\
7	15.5555555555556\\
8	16.7901234567901\\
9	19.0123456790123\\
10	22.2222222222222\\
11	22.962962962963\\
12	23.2098765432099\\
13	24.1975308641975\\
14	24.4444444444444\\
15	25.1851851851852\\
16	25.4320987654321\\
17	25.4320987654321\\
18	25.4320987654321\\
};
\label{lab:cdr_0}
\addlegendentry{Normal}


\addplot [color=mycolor5, line width=2.0pt, mark size=1.5pt, mark=*, mark options={solid, fill=mycolor5, mycolor5}]
  table[row sep=crcr]{%
1	7.27272727272727\\
2	19.0909090909091\\
3	23.6363636363636\\
4	26.3636363636364\\
5	34.5454545454545\\
6	36.3636363636364\\
7	41.8181818181818\\
8	44.5454545454545\\
9	47.2727272727273\\
10	51.8181818181818\\
11	51.8181818181818\\
12	51.8181818181818\\
13	51.8181818181818\\
14	51.8181818181818\\
15	53.6363636363636\\
16	53.6363636363636\\
17	53.6363636363636\\
18	53.6363636363636\\
};
\label{lab:cdr_half}
\addlegendentry{CDR $0.5$}

\addplot [color=mycolor15, line width=2.0pt, mark size=1.5pt, mark=*, mark options={solid, fill=mycolor15, mycolor15}]
  table[row sep=crcr]{%
1	8.16326530612245\\
2	12.2448979591837\\
3	16.3265306122449\\
4	22.4489795918367\\
5	34.6938775510204\\
6	38.7755102040816\\
7	48.9795918367347\\
8	53.0612244897959\\
9	61.2244897959184\\
10	69.3877551020408\\
11	69.3877551020408\\
12	69.3877551020408\\
13	69.3877551020408\\
14	69.3877551020408\\
15	71.4285714285714\\
16	73.469387755102\\
17	73.469387755102\\
18	73.469387755102\\
};
\label{lab:cdr_1}
\addlegendentry{CDR $1$}

\addplot [color=mycolor16, line width=2.0pt, mark size=1.5pt, mark=*, mark options={solid, fill=mycolor16, mycolor16}]
  table[row sep=crcr]{%
1	16.6666666666667\\
2	33.3333333333333\\
3	33.3333333333333\\
4	50\\
5	66.6666666666667\\
6	66.6666666666667\\
7	83.3333333333333\\
8	83.3333333333333\\
9	83.3333333333333\\
10	83.3333333333333\\
11	100\\
12	100\\
13	100\\
14	100\\
15	100\\
16	100\\
17	100\\
18	100\\
};
\label{lab:cdr_2}
\addlegendentry{CDR $2$}
\node[right, align=left]
at (axis cs:14,30) {$103/405 = 25$\%};
\node[right, align=left]
at (axis cs:14.3,59) {$59/110 = 54$\%};
\node[right, align=left]
at (axis cs:14.6,78) {$36/49 = 73$\%};
\node[right, align=left]
at (axis cs:14.9,94) {$6/6 = 100$\%};
\end{axis}
\end{tikzpicture}%
}
		\caption{\small The cumulative number of outliers detected during the iterative data-cleaning strategy with respect to the total number of patients in the training dataset from the corresponding clinical dementia rating (CDR) levels. Annotated is the amount of outliers$/$total number of patients in each CDR level. These results were obtained using the threshold parameter $R = 1.96$.\label{fig:data_trial}}
		\vspace{-3.3mm}
	\end{figure}
	\vspace{-2mm}\subsection{Dataset}
	
	Due to the lack of reference ground-truth BA labels, the validation of the proposed iterative data-cleaning strategy poses a key challenge. For this purpose, we investigate the performance of the introduced training strategy on an age-balanced subset from the OASIS-3 brain dataset \cite{Extra6}. Unlike the IXI dataset which contains MRI scans from healthy patients, the OASIS-3 dataset encompasses T1-weighted MR scans from anonymized cognitively healthy individuals as well as patients suffering from dementia due to Alzheimer's disease. The utilization of a dataset containing a mixture of healthy and dementia patients would serve to test the capability of the iterative framework in detecting outlier patients. The scans were acquired according to a flip angle of 10°, echo time of 4 ms and a repetition time of 9.7 ms. The degree of cognitive deterioration in the Alzheimer's patients is indicated by the clinical dementia rating (CDR), which distinguishes between questionable, mild and moderate dementia by the CDR scores of 0.5, 1 and 2, respectively \cite{Extra7}. 
	
	In total, we utilize a subset of 1230 MRI scans from 950 patients in the age-range of 48-97 years. To ensure that the Age-Net is trained on a class-balanced dataset, we allocate 565 MR scans from 405 healthy patients and 185 scans from 165 Alzheimer's patients for training the proposed framework. The remaining 490 scans from 380 patients (270: healthy, 110: Alzheimer's) are assigned as the test set. The histogram of the utilized OASIS-3 data is depicted Fig.~\ref{fig:oasis_dist}. The same pre-processing pipeline described previously in Sec.~\ref{sec:data-ca} was also applied with GM MRI chunks of matrix size $121 \times 145 \times 6$ being fed to the framework as inputs. Similar to the experiments in the previous section, we utilize GM content instead of raw MRI volumes and we also exclude the first and last 3 chunks from our procedure as they mainly represent skull regions.
	
	\vspace{-2mm}\subsection{Experiments}
		\begin{figure*}
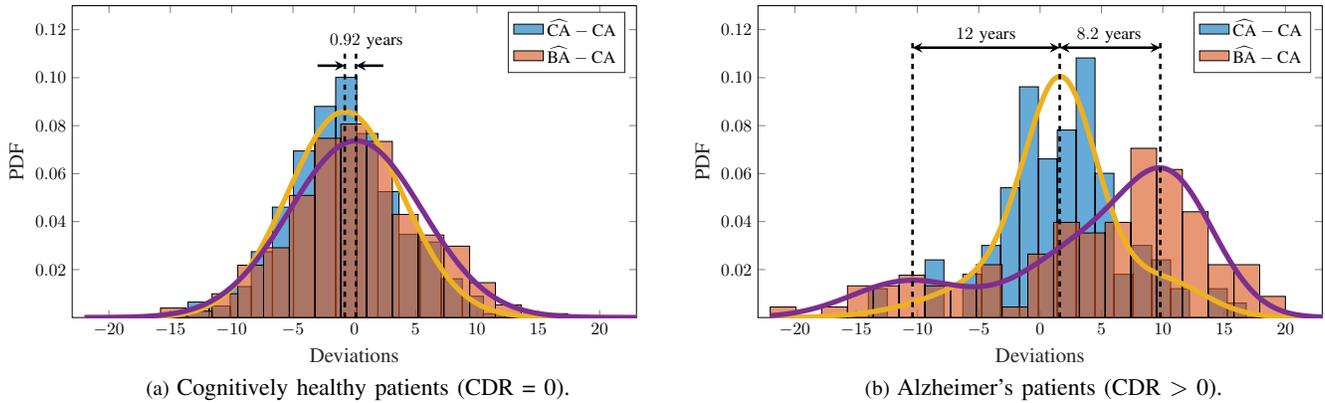

		\captionsetup[subfigure]{oneside,margin={1.cm,0cm}}
		\subfloat[\small Cognitively healthy patients (CDR = 0).\label{fig:deviations_healthy}]{\resizebox{0.95\columnwidth}{!}{\input{healthy_k_1.tex}}}
		\qquad
		\subfloat[\small Alzheimer's patients (CDR $>$ 0).\label{fig:deviations_alz}]{\resizebox{0.95\columnwidth}{!}{\input{alz_k_3.tex}}}
		\caption{\small The PDF of the deviations between the estimated ages and the ground-truth CA labels. The depicted lines (\ref{lab:healthy_ca}) and (\ref{lab:healthy_ba}) represent the best-fit distribution for the CA and BA networks, respectively. \label{fig:deviations_all}}
		\vspace{-2mm}
	\end{figure*}
	In previous studies, it has been reported that Alzheimer's disease correlates directly with abnormal brain characteristics, particularly, accelerated aging \cite{E16}. We apply this observation to evaluate the capability of the iterative data-cleaning strategy in detecting atypically-aging individuals. More specifically, we count the number of patients flagged as outliers by the proposed training strategy. Further, we analyze the percentage of cognitively healthy individuals (CDR = 0) and Alzheimer's patients (CDR = 0.5, 1, 2) detected as outliers with respect to their corresponding populations in the training dataset. We hypothesize that the proposed training strategy should be capable of accurately detecting patients with mild and moderate dementia as those theoretically exhibit the most pronounced atypical-aging characteristics. 
	
	Additionally, we compare the final predicted BAs, after applying the iterative data-cleaning framework, against the age estimates from an Age-Net trained by using the CA as ground-truth labels. We illustrate, and subsequently analyze, the distributions of the resultant age estimates of the two frameworks. This was conducted separately for both cognitively healthy and Alzheimer's patients. Additional experiments examining the effects of data distribution and balancing on the performance of the iterative strategy are presented in Appendix~C.

	Motivated by the recent interest in providing explainable DL frameworks, we attempt to shed light onto the reason beyond the predicted BA decision of our proposed network. For this purpose, we employ state-of-the-art DL visualization techniques to highlight the most significant brain regions accounting to the network's prediction from patients labeled as outliers or healthy by the iterative strategy. Specifically, we utilize at inference a combination of saliency-map visualizations \cite{Extra3,Extra4} together with GradCAM++ \cite{Extra5} for more fine-grained visualization maps. 
	
	Class activation mapping (CAM) based techniques, such as GradCAM++, are suitable for object detection and classification tasks from a group of objects in an image. However, they tend to generally produce coarse-grained visualizations that may not accurately reflect the relevant brain regions. In contrast, guided backpropagation techniques produce more fine-grained visualization maps. This comes with the cost of demanding considerable memory and computational overheads as it requires computing and backpropagating gradients throughout the network. Recently in \cite{Extra4}, authors proposed a gradient-free approach to compute fine-grained visualization that can be combined with coarse-grained techniques, such as GradCAM++, to efficiently produce refined visualizations. The resultant outputs of each technique are combined via a product operation to obtain the final visualization maps. We compare the differences between the visualization maps from healthy and Alzheimer's patients in the same age groups. Also, we analyze the visualizations of healthy individuals (CDR = 0) who were deemed by the framework as exhibiting atypical-aging characteristics, thus flagged as outliers. 
	
	It is important to note that the results of these visualization techniques do not imply functional brain activation such, e.g., neuronal activity. Rather, it is utilized as a means to explain the network's decisions and highlight the differences between the results of healthy and outlier patients. 
	
	\section{Results and Discussion}\label{sec:results}
	\begin{figure*}[!t]
		\centering
		\includegraphics[width=0.95\textwidth]{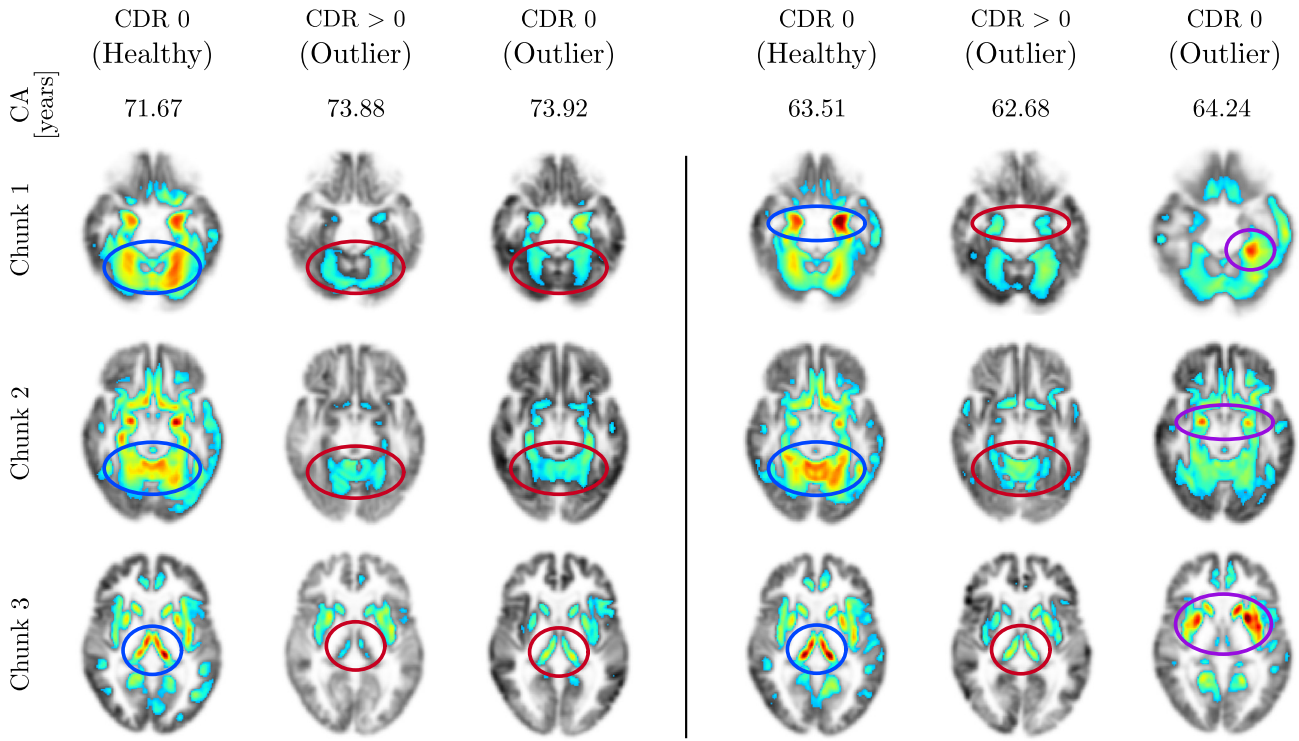}
		\caption{\small Examples of the brain visualization maps for 3 axial chunks from healthy and outlier patients as detected by the proposed iterative strategy. Each column represents scans from a patient of the annotated CA value. The prominent activations in healthy patients were highlighted by (\myelipseblue) whereas the corresponding activations in Alzheimer's patients were labeled via (\myelipsered). Uncommon or unique activations detected in cognitively healthy outliers were labeled by (\myelipsepurple). \label{fig:visualization}}
		\vspace{-2mm}
	\end{figure*}
	
	The first step towards analyzing the proposed iterative data-cleaning strategy is to examine the detected outliers across consecutive iterations. As depicted in Fig.~\ref{fig:data_trial}, a total of 18 training iterations were conducted before termination. This is due to satisfying the pre-defined stopping condition with no outliers detected in three successive iterations. Upon examining the population of patients with moderate dementia (CDR = 2, \ref{lab:cdr_2}), it is observed that the proposed strategy detects all aforementioned patients after 11 iterations. For Alzheimer's patients with mild dementia (\ref{lab:cdr_1}), 36 out of 49 patients were flagged as outliers in 16 iterations, amounting to a total of 73\% of the CDR 1 training population. For questionable dementia (CDR = 0.5, \ref{lab:cdr_half}), 54\% of this population were detected as outliers in 15 iterations. Conversely, for cognitively healthy individuals (\ref{lab:cdr_0}) a substantially smaller percentage of patients (25\%) were flagged as outliers. Compared to the number of outliers detected from Alzheimer's patients, it is realistic for cognitively healthy individuals to less frequently exhibit atypically-aging characteristics. The above findings indicate that this training algorithm is capable of detecting atypical-aging characteristics whether Alzheimer's related or not. This demonstrates that the proposed training strategy is not only restricted to Alzheimer's detection but reveals a higher generalizable potential. For instance, it can potentially be extended to different sources of brain deterioration such as tumours and lesions. Subsequently, all flagged outliers are removed from the training samples to create a typical-aging dataset ($\textrm{BA} \approx \textrm{CA}$). Thereafter, an Age-Net framework is trained for the task of BA estimation using this filtered dataset. The distribution of the remaining training dataset is also depicted in Fig.~\ref{fig:oasis_dist}.
	
	Additionally, we also compare the predicted ages of the test dataset from the proposed BA estimation framework (after iterative data-cleaning) against those from a conventionally trained CA estimation Age-Net. The probability distribution functions (PDFs) of the deviations between the predicted ages and ground-truth CA labels for both frameworks are presented in Fig.~\ref{fig:deviations_all}. For the cognitively healthy population, minor perceivable differences can be observed in the distributions of both frameworks. This is illustrated in Fig.~\ref{fig:deviations_healthy} with both of them adopting a normal distribution with a mean-deviation shift of 0.92 years. However, in Fig.~\ref{fig:deviations_alz} the predicted BA ages for Alzheimer's patients manifest a substantially different behavior with a multimodal distribution (\ref{lab:alz_ba}) as opposed to a normal distribution (\ref{lab:alz_ca}) by the CA framework. Since Alzheimer's disease was found to correspond with abnormal brain aging \cite{E16}, it follows that the predicted BA scores for patients with dementia must exhibit a significant deviation from the corresponding CA labels. This desired behavior is signified in the proposed BA estimation framework with the majority of the predicted ages exhibiting an over-aging of approximately 8.2 years compared to the CA framework. We hypothesize that this reflects the capability of the framework in recognizing true BA behavior. Also, a relatively smaller population of Alzheimer's patients reveals under-aging behavior. 
	
 	For the final set of experimental evaluations, we analyze the visualization maps from patients identified by the data-cleaning strategy as exhibiting atypical-aging characteristics. Thus, these patients were removed from the final BA training dataset. It is important to distinguish that these visualization maps reflect the network predictions rather than the actual brain voxel intensities. We also compare the resultant outlier visualization maps against those extracted from healthy individuals. As shown in Fig.~\ref{fig:visualization}, cognitively healthy individuals exhibit strong activations in the amygdala, hippocampus and thalamus as indicated by (\myelipseblue). This agrees with prior work investigating voxel-based morphometry of brain MRI of healthy patients \cite{Extra9}. On the other hand, patients suffering from dementia (CDR $>$ 0) identified as outliers manifest relatively weaker activations in the same regions, as indicated\linebreak by (\myelipsered). However, it is interesting to point out that cognitively healthy patients flagged as outliers (CDR = 0) reveal a noteworthy behavior. In some instances, the visualization maps from the aforementioned patients closely resemble the maps from dementia patients with similar highlighted regions of low intensity. For other outlier patients of CDR 0, the resultant visualization maps exhibit unconventional behavior with different highlighted regions of high intensity compared to cognitively healthy patients. For example, the regions in and around the fusiform gyrus depict high activations in the outlier patients, as specified by (\myelipsepurple), in contrast to healthy individuals. On the whole, Alzheimer's patients generally exhibit weakened networks activations compared to healthy individuals. Also, the CDR 0 outlier patients are either similar to dementia patients or display unique activations. We are of the opinion that the study of the activated brain regions for outliers could potentially prove to be beneficial for radiologists as it may assist in the early detection of disorders along with other various applications. 
 	
 	This initial study reveals the potential of utilizing MRI scans for BA estimation. The concept of incorporating an iterative training algorithm for approximating brain BA labels shows merit by detecting the majority of patients with moderate and mild dementia as outliers. This does not come at the expense of an overt detection of cognitively healthy patients as atypically-aging. Additionally, a visualization study highlighted the possibility of utilizing the proposed BA estimation framework to discover the deviation of seemingly healthy patients from their respective age groups. 
 	
 	As this study is among the first of its kind, our work has raised more questions than it has provided answers. Further clinical assessments by radiologists are necessary for the validation of the introduced framework. More specifically, a key question is the applicability of the BA framework on expanded MRI datasets from various organ systems. This could be a step towards achieving an organ-specific BA assessment for patients using whole-body MRI scans. This serves to evaluate the accelerated or delayed-aging characteristics of various organs within patients. Additionally, we plan to apply the proposed BA pipeline on different brain MRI datasets with different disorders associated with abnormal aging such as epilepsy, schizophrenia and HIV \cite{Resub-2,Resub-3,Resub-4}. The prior experiments must also include a longitudinal analysis to observe the differences in the predicted BA throughout different stages of diseases. The correlation between the deviations of the estimated BA scores and the presence of different disorders should be studied by radiologists together with the resultant visualization maps. Also, in the future, we plan to investigate the possibility of utilizing the BA visualization maps for anomaly detection of lesions and disorders in an unsupervised setting. Furthermore, an in-depth assessment must be conducted to investigate the reason beyond the differences between the BA and conventional CA predictions from a clinical perspective.
 	
 	\begin{table}[!b]
 		\caption{\small \\Quantitative comparison for CA estimation on OASIS-3 dataset.\label{tab:results-CA-oasis}}
 		\centering
 		\setlength\arrayrulewidth{0.05pt}
 		\small
 		\bgroup
 		\midsepremove
 		\def\arraystretch{1.45}
 		\resizebox{\columnwidth}{!}{%
 			\begin{tabular}{l ccccc}
 				\toprule
 				Model & MAE & SD & Bias & RMSE & Corr.\\ 
 				\midrule  
 				\myrowcolour 3D-Peng \cite{Resub-5} & 4.170 & 5.333 & 0.663 & 5.374 & 0.769\\
 				Age-Net-Gender & \textbf{3.610} & \textbf{4.754} & \textbf{0.218} & \textbf{4.759} & \textbf{0.860} \\
 				\bottomrule
 			\end{tabular}
 		}
 		\egroup
 	\end{table}

	\section{Conclusion}\label{sec:conc}
	
	In this work, we present an initial study for organ-specific BA estimation using MRI scans. As an initial step, we develop a CA estimation framework capable of outperforming the current state-of-the-art MRI-based regression networks. Furthermore, we introduce a novel iterative training algorithm for excluding outlier patients exhibiting atypical-aging characteristics. This leads to the creation of a reference dataset where the available CA labels reflect the BA behavior. Upon validating the proposed BA framework on an Alzheimer's dataset, the majority of patients with mild or moderate dementia were accurately detected as outliers. Moreover, the framework was found to be effective in detecting accelerated aging in Alzheimer's patients in comparison to conventional CA estimation. Finally, an analysis was performed using established DL visualization techniques which reflects the potential of the introduced framework in discovering deviations of seemingly healthy patients from their respective age groups. In the future, we plan to expand the framework via the utilization of recent advances in Bayesian neural networks and uncertainty detection techniques to enhance the outlier detection procedure \cite{Resub-7,Resub-8}. Also, an extension to different organ systems with whole-body MRI data will be investigated.
	
	\begin{table}[!t]
		\caption{\small \\Number of detected outliers from different CDR groups in response to different outlier detection thresholds. All CDR 2 patients reported a 100\% detection rate for all investigated thresholds.\label{tab:Rs}}
		\centering
		\setlength\arrayrulewidth{0.25pt}
		\large
		\bgroup
		\midsepremove
		\def\arraystretch{1.45}
		\setlength\heavyrulewidth{0.21ex}
		\setlength\lightrulewidth{0.13ex}
		\resizebox{\columnwidth}{!}{%
			\begin{tabular}{l ccc}
				\toprule
				Outliers [\%] & CDR $= 0$ & CDR $= 0.5$ & CDR $= 1$\\ 
				\midrule  
				\myrowcolour $R = 1$ & $148/405 = 37 \%$ & $67/110 = 61 \%$ & $37/49 = 76 \%$\\
				$R = 2$ & $103/405 = 25 \%$ & $59/110 = 54 \%$ & $36/49 = 73 \%$\\
				\myrowcolour $R = 3$ & $77/405 = 19 \%$ & $54/110 = 49 \%$ & $32/49 = 65 \%$\\
				$\gamma_{\textrm{fixed}}$ & $252/405 = 62 \%$ & $77/110 = 70 \%$ & $40/49 = 82 \%$\\
				\bottomrule
			\end{tabular}
		}
		\egroup
	\end{table}
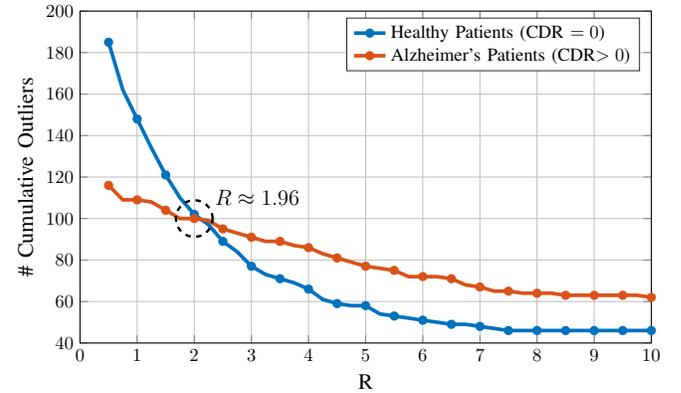
\begin{figure}[!t]
	\resizebox{0.48\textwidth}{!}{\begin{tikzpicture}

\begin{axis}[%
width=4.3in,
height=2.5in,
at={(0in,0in)},
scale only axis,
xmin=0,
xmax=10,
ymin=40,
ymax=200,
ylabel={\large \# Cumulative Outliers},
xlabel={\large R},
axis background/.style={fill=white},
xmajorgrids,
ymajorgrids,
legend style={legend cell align=left, align=left, draw=white!15!black}
]
\addplot [color=mycolor4, line width=2.0pt, mark size=1.5pt, mark=*, mark options={solid, fill=mycolor4, mycolor4}, mark indices = {2,4,6,8,10,12,14,16,18,20,22,24,26,28,30,32,34,36,38,40}]
table[row sep=crcr]{%
0.25	190\\
0.5	185\\
0.75	162\\
1	148\\
1.25	134\\
1.5	121\\
1.75	110\\
2	102\\
2.25	97\\
2.5	89\\
2.75	84\\
3	77\\
3.25	73\\
3.5	71\\
3.75	69\\
4	66\\
4.25	61\\
4.5	59\\
4.75	58\\
5	58\\
5.25	54\\
5.5	53\\
5.75	52\\
6	51\\
6.25	50\\
6.5	49\\
6.75	49\\
7	48\\
7.25	47\\
7.5	46\\
7.75	46\\
8	46\\
8.25	46\\
8.5	46\\
8.75	46\\
9	46\\
9.25	46\\
9.5	46\\
9.75	46\\
10	46\\
};
\addlegendentry{Healthy Patients (CDR $=0$)}
\addplot [color=white, line width=2.6pt,forget plot]
table[row sep=crcr]{%
	0.2	191\\
	0.25	190\\
	0.5	185\\
};
\addplot [color=mycolor5, line width=2.0pt, mark size=1.5pt, mark=*, mark options={solid, fill=mycolor5, mycolor5}, mark indices = {2,4,6,8,10,12,14,16,18,20,22,24,26,28,30,32,34,36,38,40}]
table[row sep=crcr]{%
	0.25 117\\
0.5	116\\
0.75	109\\
1	109\\
1.25	108\\
1.5	104\\
1.75	100\\
2	100\\
2.25	99\\
2.5	95\\
2.75	93\\
3	91\\
3.25	89\\
3.5	89\\
3.75	87\\
4	86\\
4.25	83\\
4.5	81\\
4.75	79\\
5	77\\
5.25	76\\
5.5	75\\
5.75	72\\
6	72\\
6.25	72\\
6.5	71\\
6.75	68\\
7	67\\
7.25	65\\
7.5	65\\
7.75	64\\
8	64\\
8.25	64\\
8.5	63\\
8.75	63\\
9	63\\
9.25	63\\
9.5	63\\
9.75	63\\
10	62\\
};
\addlegendentry{Alzheimer's Patients (CDR$>0$)}
\addplot [color=white, line width=2.6pt,forget plot]
table[row sep=crcr]{%
	0.2	116\\
	0.25	117\\
	0.5	116\\
};
\draw (axis cs:2,100)[dashed,line width=1.3pt] ellipse (0.35cm and 0.35cm);
\node[font=\fontsize{12}{0}\selectfont,right, align=left]
at (axis cs:2.25,110) {\large $R\approx 1.96$};
\end{axis}
\end{tikzpicture}%
}
	\caption{\small The cumulative number of detected outliers for both healthy and Alzheimer's patients over different values for the threshold parameter $R$.\label{fig:r_exp}}
		\vspace{-2mm}
\end{figure}
	
\section*{Appendix}

In the following sections, additional experimental evaluations are conducted to investigate and highlight the performance of several aspects of the proposed pipeline.

\subsection{CA Comparative Analysis on OASIS-3 Dataset \label{sec:appendixA}}

To investigate the performance of the proposed Age-Net architecture on a dataset containing both healthy and patients with neurological disorders, we extended the quantitative comparative analysis, previously presented in Table~\ref{tab:results}. 
More specifically, we examine the behavior of the best two performing CA estimation architectures, i.e., 3D-Peng and Age-Net-Gender, on the OASIS-3 brain dataset. For training and testing, we utilize the same pre-processing and data-splitting presented previously in Sec.~\ref{sec:data-ca}. Both models were trained using the same hyper-parameters till convergence. The quantitative results are presented in Table~\ref{tab:results-CA-oasis}. Due to training on a mixture of healthy and Alzheimer's patients, the performance of both frameworks is quantitatively worse when compared to training solely on the healthy population in the IXI dataset. However, the proposed Age-Net-Gender approach still quantitatively outperforms the 3D-Peng framework by a MAE of 0.5 years.

\begin{figure*}[t!]
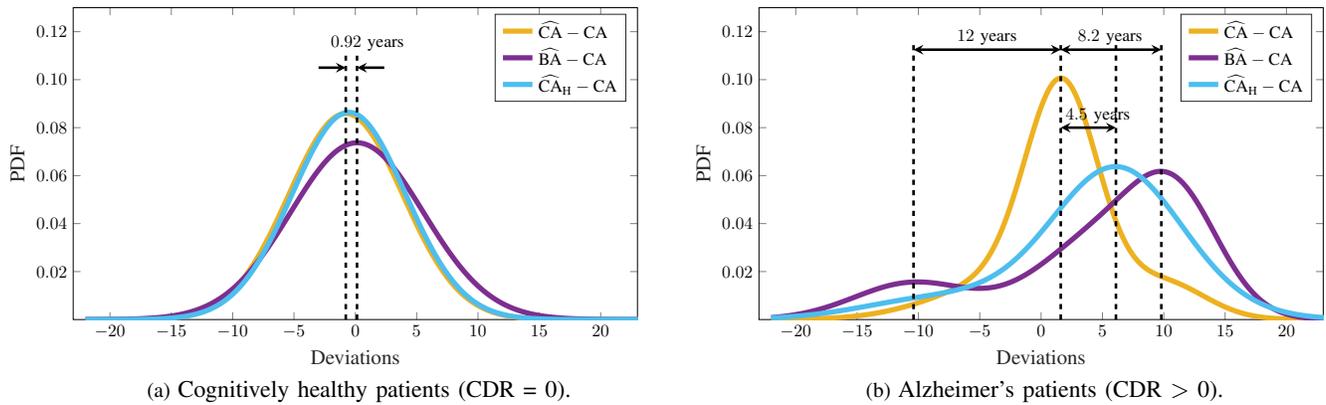

	\captionsetup[subfigure]{oneside,margin={1.cm,0cm}}
	\subfloat[\small Cognitively healthy patients (CDR = 0).\label{fig:deviations_healthy_exp_healthy_only}]{\resizebox{0.95\columnwidth}{!}{\input{exp_healthy_only_healthy.tex}}}
	\qquad
	\subfloat[\small Alzheimer's patients (CDR $>$ 0).\label{fig:deviations_alz_exp_healthy_only}]{\resizebox{0.95\columnwidth}{!}{\input{exp_healthy_only_Alz.tex}}}
	\caption{\small The PDF of the deviations between the estimated ages and the ground-truth CA labels. The depicted lines (\ref{lab:alz_ca_exp_alz_more}) and  (\ref{lab:alz_ba_healthy_dominant}) represent the best-fit distribution for the CA and BA networks trained on a mixed healthy and Alzheimer's dataset, respectively, while the line (\ref{lab:alz_ba_Alz_dominant}) represents the CA framework trained solely using healthy population ($\widehat{\textrm{CA}}_{\textrm{\tiny H}}$). \label{fig:deviations_all_exp_healthy_only}}
		\vspace{-2mm}
\end{figure*}

\subsection{Investigation of Different Outlier Detection Thresholds}

The outlier detection threshold presented in Eq.~\ref{eq:th} is an integral component in our proposed BA estimation pipeline. In the above manuscript, we have set the value for the threshold parameter $R$ to 1.96. This value was chosen empirically upon experimenting with different values of $R$. As shown in Fig.~\ref{fig:r_exp}, as $R$ increases this, in turn, is reflected as a subsequent decrease in the number of detected outliers in both healthy and Alzheimer's patients. However, at $R\approx1.96$ the number of healthy and Alzheimer's outliers intersect. As such, this intersection point was chosen empirically for our proposed iterative pipeline.

From another perspective, the currently utilized outlier detection threshold is based on the chunk data-feeding strategy. It does not translate directly to dealing with full MRI volumes as in this case there exists a single age estimate for each patient. Thus, no patient-dependent uncertainty value can be utilized as a formulation for the outlier detection threshold. To investigate training the BA pipeline with full MRI volumes, we examine the performance of a new patient-independent threshold calculated as: 
\begin{equation}
	\gamma_{\textrm{fixed}} = \frac{1}{N} \sum_{n=1}^{N} \left| \widehat{\textrm{CA}}_n - \textrm{CA}_n   \right|
\end{equation}
This absolute CA deviation $D_n = |\widehat{\textrm{CA}}_n - \textrm{CA}_n| $ for each patient $n$ is then compared against the above threshold. An outlier is detected only if $D_n > \gamma_{\textrm{fixed}}$. The number of detected outliers for this whole-volume based approach is compared against different $R$ values for the proposed chunk-based pipeline in Table~\ref{tab:Rs}. The number of detected outliers illustrate that this fixed threshold results in an overt detection of healthy patients. We hypothesize that this degradation in performance is due to not utilizing the patient-dependent uncertainty values which we have identified as an integral component in our framework.

In a subsequent work, dealing with full MRI volumes instead of chunks can be tackled by incorporating the concepts of aleatoric or epistemic uncertainties. For instance, in case of epistemic uncertainty, the same MRI volume can be fed multiple times to the network with each iteration producing a different CA estimate due to having weights and biases defined as distributions (Bayesian neural network) rather than deterministic values \cite{Resub-7}. In this manner, patient-dependent uncertainty can still be calculated using the standard deviation over the different CA estimations for each patient. With regards to aleatoric uncertainty, the network is trained to minimize the evidence lower bound (ELBO) loss function to produce a predicted age estimate together with an aleatoric uncertainty value for each patient, representing the noise inherent in the input MRI volume \cite{Resub-8}. This can be later adopted for the outlier detection threshold. We plan to experiment with both types of uncertainty estimation in further future work.

\subsection{Population Distributions Effect on BA Pipeline}

\subsubsection{Majority Healthy Training Population}

To highlight the performance of the proposed BA estimation pipeline, we train a new Age-Net architecture for CA estimation solely using MRI scans of healthy individuals (CDR 0) within the OASIS-3 dataset. We compare the resultant CA scores against the proposed BA and CA trained on a mixed dataset. The PDF of deviations between the estimated ages of the test dataset and their CA labels are depicted in Fig.~\ref{fig:deviations_all_exp_healthy_only}. Based on the results for cognitively healthy individuals in Fig.~\ref{fig:deviations_healthy_exp_healthy_only}, the model trained with only healthy individuals is giving the same performance as the CA model trained with a mixed population, while the BA model is showing a marginal shift in the deviations. Moreover, in the case of Alzheimer's patients, the model trained only on CDR 0 patients results in a less pronounced over-aging behavior of 4.5 years, as illustrated by Fig.~\ref{fig:deviations_alz_exp_healthy_only}. This is in comparison to 8.2 years deviations between the proposed BA and CA frameworks trained on the mixed dataset. We hypothesis that this is because our proposed model removes a portion of the CDR 0 patients as outliers. This helps to emphasize the over-aging characteristics of the outlier patients.

\begin{figure*}[t!]
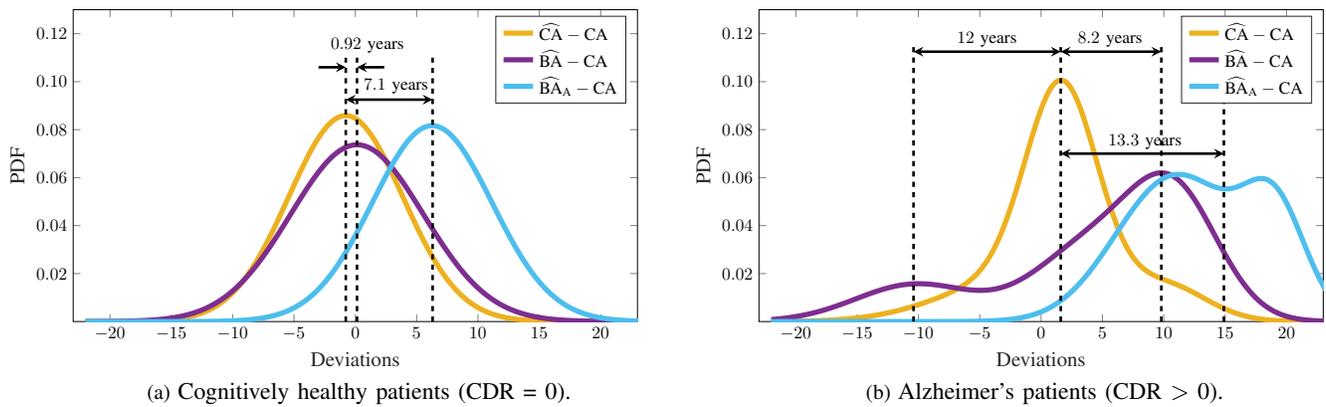

	\captionsetup[subfigure]{oneside,margin={1.cm,0cm}}
	\subfloat[\small Cognitively healthy patients (CDR = 0).\label{fig:deviations_healthy_exp_alz_more}]{\resizebox{0.95\columnwidth}{!}{\input{exp_alz_more_healthy1.tex}}}
	\qquad
	\subfloat[\small Alzheimer's patients (CDR $>$ 0).\label{fig:deviations_alz_exp_alz_more}]{\resizebox{0.95\columnwidth}{!}{\input{exp_alz_more_Alz1.tex}}}
	\caption{\small The PDF of the deviations between the estimated ages and the ground-truth CA labels. The depicted lines (\ref{lab:alz_ca_exp_alz_more}) and  (\ref{lab:alz_ba_healthy_dominant}) represent the best-fit distribution for the CA and BA networks trained on a mixed healthy and Alzheimer's dataset, respectively, while the line (\ref{lab:alz_ba_Alz_dominant}) represents the BA framework trained using a majority of Alzheimer's population ($\widehat{\textrm{BA}}_{\textrm{\tiny A}}$). \label{fig:deviations_all_exp_alz_more}}
		\vspace{-4mm}
\end{figure*}

\vspace{2mm}\subsubsection{Majority Alzheimer's Training Population}

In previous experiments, we assume that the majority of the training population is healthy with no neurological disorders. We also discussed previously that that absence of a diagnostic disease does not guarantee that the addressed population exhibit typical aging characteristics. However, we still assume that the majority of CDR 0 patients are typically-aging. In this manner, our framework attempts to learn the true aging characteristics of the dominant population in the training dataset while detecting the minority populations as outliers. Disturbing the aforementioned assumption has severe consequences on the stability and performance of the training procedure. 

To investigate this further, we conduct a BA estimation experiment where the Alzheimer's patients constitute a majority 70\% of the training population while the healthy patients represent only 30\% of the training data. In this experiment, since the healthy training population is now the minority group, they are more significantly detected as outliers amounting to a total of 63\% of the CDR 0 patients. Moreover, Alzheimer's patients exhibit no consistent correlation between their aging characteristics. As such, the model is not capable of recognizing a common aging behavior for these majority patients. This leads to 70\% of their respective training population to be detected as outliers. This leaves only 32\% of the training dataset available to train the final BA regression network.

Upon examining the results of this final model in comparison to the traditional CA framework and the BA results (trained on a healthy majority population), several observations can be made. This is pointed out in Fig.~\ref{fig:deviations_all_exp_alz_more} which depicts the PDF of deviations between the estimated ages of the test dataset and their CA labels for the aforementioned models. The healthy patients in Fig.~\ref{fig:deviations_healthy_exp_alz_more} are falsely classified by the unbalanced model as atypically-aging. This reinforces that the iterative strategy detects the minority patients as outliers. Also, Alzheimer's patients are noted as over-aging due to the majority of them being detected as outliers in the training dataset, as shown in Fig.~\ref{fig:deviations_alz_exp_alz_more}. As seen by the above results, unbalancing the training dataset significantly affects the final predicted BA of the healthy and typically-aging population. To conclude, the training database should be primarily made up of a majority of relatively healthy subject with no significant disorders in order for the model to learn typical aging characteristics.

\bibliographystyle{IEEEtran}
	

\bibliographystyle{IEEEtran}

\end{document}